\documentclass[fleqn,twoside]{article}

\usepackage{epsfig}
\usepackage{graphicx}
\usepackage{rotating}
\usepackage{fleqn}
\usepackage{amsmath}
\usepackage{espcrc2}

\newcommand{\pvec}{{\bf p}}
\def\etal {{\it et~al.,}}
\input babarsym

\title{\begin{flushleft}\small{Presented at QCD 04: High Energy Physics International Conference
       in Quantum Chromodynamics, Montpellier, France, 5-9 Jul 2004.}\\
       \vskip 5pt
       \small{SLAC-PUB-10773}
       \end{flushleft}         
       \vskip 20pt
       Selected Topics in \CP\ Violation and Weak Decays from \babar}

\author{J.J.Back\address{Department of Physics, University of Warwick, \\
	Coventry, CV4 7AL, UK
	(on behalf of the \babar\ Collaboration)}%
	\thanks{Work supported in part by the Department of Energy
	contract DE-AC02-76SF00515.}}

\begin{document}

\begin{abstract}
We present branching fraction and \CP\ asymmetry 
results for a variety of charmless
$B$ decays based on up to 124~\invfb collected by 
the \babar\ experiment running near the \FourS resonance at 
the \pep2 \epem $B$-factory.
\end{abstract}

\maketitle

\section{Introduction}

\CP\ violation has been established in the $B$-meson system~\cite{BabarCPV}~\cite{BelleCPV} and
$B$-factories are now focusing their attention on over-constraining the
angles and sides of the Unitarity Triangle, which is a partial representation of the
Cabibbo-Kobayashi-Maskawa (CKM) matrix~\cite{CKM}. The study of charmless $B$ decays
allows us to make such measurements and also to probe physics beyond the Standard
Model (SM). In this paper, we present the preliminary results of a few charmless analyses.

\section{\boldmath{\CP} Asymmetries in $B$ Decays}

For charged $B$ decays, \CP\ violation can occur when
we have at least two interfering amplitudes that have different weak and
strong phases. This is known as direct \CP\ violation, and manifests itself 
as an asymmetry in the partial decay rates for particle and anti-particle:
\begin{equation}
{\cal{A}}_{\rm{direct}} = \frac{\Gamma (B^- \ra f^-) - \Gamma(B^+ \ra f^+)}
{\Gamma (B^- \ra f^-) + \Gamma(B^+ \ra f^+)}
\end{equation}
where $\Gamma(B^- \ra f^-)$ is the decay rate
for $B^- \ra f^-$, and $\Gamma(B^+ \ra f^+)$ is the decay
rate for the charge-conjugate process.

For neutral $B$ decays, \CP\ violation is present when we
have interference between $B^0$ and $\bar{B}^0$ decays,
with and without mixing, and manifests
itself as a difference in the decay rates of the $B$ mesons to a common
final state.
The asymmetric beam configuration of the \babar\ experiment
provides a boost of $\beta \gamma = 0.56$ to the \FourS\
in the laboratory frame, which allows the measurement of the
decay time difference $\Delta t$ between the $B^0$ and $\bar{B}^0$ mesons along
the beam axis. We fully reconstruct the signal $B$ decay
and partially reconstruct the other $B$ meson in order
to determine its flavour, i.e. whether it is $B^0$ or $\bar{B}^0$. We
can then measure the \CP-violating parameters $C$ and $S$ by fitting
the following function to the decay time distribution (taking
into account experimental resolution effects):
\begin{eqnarray}
f(\Delta t) & = & \frac{e^{-|\Delta t|/\tau}}{4\tau} 
[1 + Q_{\rm{tag}} S~{\rm sin}(\Delta m_d \Delta t) \nonumber \\
& & - Q_{\rm{tag}} C~{\rm cos}(\Delta m_d \Delta t)],
\end{eqnarray}
where $Q_{\rm{tag}} = 1(-1)$ when the tagging meson is a $B^0$ ($\bar{B}^0$),
$\tau$ is the mean $B^0$ lifetime, and $\Delta m_d$ is the $B^0$-$\bar{B}^0$
oscillation frequency corresponding to the mass difference between the two mesons.
The presence of mixing-induced \CP\ violation would give a non-zero value
for $S$, while direct \CP\ violation would be indicated by a non-zero value of $C$.

\section{The \babar\ Detector}

The results presented in this paper are based on an integrated luminosity
of up to 124~\invfb collected at the \FourS\ resonance with the 
\babar\ detector~\cite{babardet}
at the \pep2\ asymmetric \epem\ collider at the Stanford Linear Accelerator
Center. Charged particle track parameters are measured by a 
five-layer double-sided silicon vertex tracker and a 40-layer drift chamber
located in a 1.5-T magnetic field. Charged particle identification is
achieved with an internally reflecting ring imaging Cherenkov detector
and from the average \dedx energy loss measured in the tracking devices. Photons
and neutral pions ($\piz$s) are detected with an electromagnetic calorimeter (EMC) consisting
of 6580 CsI(Tl) crystals. An instrumented flux return (IFR), containing multiple
layers of resistive plate chambers, provides muon and long-lived hadron
identification.

\section{\boldmath{$B$} Decay Reconstruction}
\label{sec:BDecayReco}

\begin{figure}[hbt!]
\begin{center}
\epsfig{file=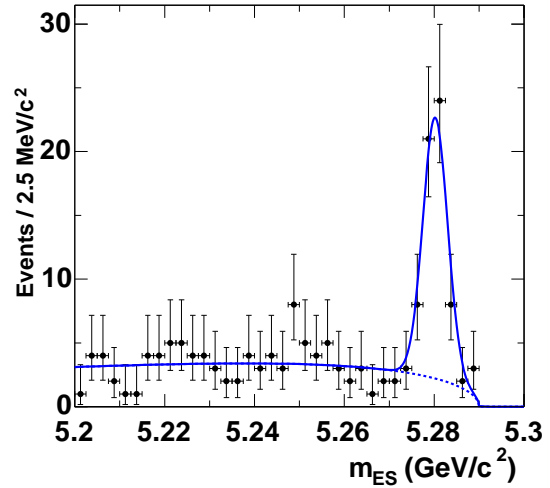, clip=, width=0.45\textwidth}
\caption{Distribution of the $\mes$ variable for $B^0 \ra \phi K^{*0}$ decays. 
The solid line represents the fit to all of the data, while the dotted line shows
the background contribution.}
\label{fig:BPhiKsmES}
\end{center}
\end{figure}

The $B$ meson candidates are identified kinematically using
two independent variables. The first is $\Delta E = E^{*} - E^{*}_{beam}$, which is
peaked at zero for signal events, since the energy of the $B$ candidate in the
\FourS\ rest frame, $E^{*}$, must be equal to the energy of the beam, $E^{*}_{beam}$,
by energy conservation.
The second is the beam-energy substituted mass,
$\mes = \sqrt{(E^{*2}_{beam} - \pvec^{*2}_B)}$, where $\pvec^{*}_B$
is the momentum of the $B$ meson in the \FourS\ rest frame, and must be close to
the nominal $B$ mass~\cite{pdg}. 
The resolution of $\mes$ is dominated by the beam energy
spread and is approximately 2.5~\mevcc. Figure~\ref{fig:BPhiKsmES} shows
an example $\mes$ distribution.

Several of the $B$ modes presented here have decays that involve 
$K^0_S$ and $\piz$ particles. $K^0_S$ candidates
are made by combining oppositely charged pions with requirements
made on the invariant mass (to be, typically, within 15~\mevcc of the nominal mass~\cite{pdg}),
the flight direction and decay vertex. Neutral pion candidates are
formed by combining pairs of photons in the EMC, with requirements made on the
energies of the photons and the mass and energy of the $\pi^0$. 

Significant backgrounds from light quark-antiquark continuum events
are suppressed using various event 
shape variables which exploit the difference in the event topologies
in the centre-of-mass frame between
background events, which have a di-jet structure, and signal events, 
which tend to be rather spherical. One example
is the cosine of the angle $\theta^*_T$ between the thrust axis of the signal 
$B$ candidate and the thrust axis of the 
rest of the tracks and neutrals in the event.
This variable is strongly peaked at unity for 
continuum backgrounds and has a flat distribution for signal.

Further suppression of backgrounds can be achieved using
a Fisher discriminant, which is a linear 
combination of event shape variables, such as the Legendre moments 
$L_j = \sum_{i} p_i \times |\rm{cos}\theta_i|^j$, where $\theta_i$ is the angle
with respect to the $B$ thrust axis of the track or neutral cluster $i$, $p_i$
is its momentum, and the sum excludes the signal $B$ candidate. The coefficients
of the Fisher discriminant are chosen such that the separation between signal
and background is maximised.

Sidebands in on-resonance (\FourS) data are used to characterise the light quark 
background
in $\Delta E$ and $\mes$, as well as data taken at 40~\mev below the \FourS\
resonance (``off-resonance''). The phenomenologically motivated Argus 
function~\cite{Argus} is used to fit the background $\mes$ distributions. 
Control samples are used to compare
the performance between Monte Carlo simulated events and on-resonance data.

All of the analyses have been performed ``blind'', meaning that
the signal region is looked at only after the selection criteria
have been finalised (in order to reduce the risk of bias).
Charge conjugate modes are implied
throughout this paper.

\section{\boldmath{$b \ra s \bar{s} s$} Gluonic Penguin Modes}
\label{sec:PenguinModes}

Here we describe the \CP\ violation results of the $B$-decay modes
$\phi K^0$, $K^+K^-K^0_S$, $K K^0_S K^0_S$ and $f_0(980) K^0_S$, which
are dominated by penguin-type Feynman diagrams. Neglecting CKM-suppressed
amplitudes, these decays have the same weak phase as the decay
$B^0 \ra J/\psi K^0$~\cite{PenguinRef}. However, the presence of heavy particles
in the penguin loops may give rise to other \CP-violating phases.
It is therefore important to measure \CP\ violation in these
modes and compare with the SM predictions to test whether there is physics
beyond the SM. The measured branching fractions and \CP-asymmetry parameters
are shown in Table~\ref{tab:PenguinResults}.

\begin{table*}[!htb]
\begin{center}
\caption{Measurements of the branching fractions (${\cal{B}}$) and \CP-asymmetry 
parameters for various gluonic $b \ra s\bar{s}s$ penguin modes. 
Results in square brackets denote limits at the 90\% confidence level. The first
and second uncertainties show the statistical and systematic errors, respectively.
The third error for the branching fraction of $f_0(980)K^0_S$ represents
model-dependent uncertainties.}
\label{tab:PenguinResults}
{\footnotesize
\begin{tabular}{lcccc}
\hline
Mode & ${\cal{B}} (\times 10^{-6})$ & $S$ & $C$ & ${\cal{A}}_{\rm direct}$ \\
\hline
$\phi K^0$ & --- & $0.47 \pm 0.34^{+0.08}_{-0.06}$ & $0.01 \pm 0.33 \pm 0.10$ & --- \\
$\phi K^0_S$ & --- & $0.45 \pm 0.43$ & $-0.38 \pm 0.37$ & --- \\
$K^+ K^- K^0_S$ & $(23.8 \pm 2.0 \pm 1.6)$ & $-0.56 \pm 0.25 \pm 0.04$ & $-0.10 \pm 0.19 \pm 0.09$ & --- \\
$K^+ K^0_S K^0_S$ & $(10.7 \pm 1.2 \pm 1.0)$ & --- & --- & $-0.04 \pm 0.11 \pm 0.02$ $[-0.23, 0.15]$\\
$f_0(980)K^0_S$ & $(6.0 \pm 0.9 \pm 0.6 \pm 1.2)$ & $-1.62^{+0.56}_{-0.15} \pm 0.10$ &
$0.27 \pm 0.36 \pm 0.12$ & --- \\
\hline
\end{tabular}
}
\end{center}
\end{table*}

\subsection{\boldmath{$B^0 \ra \phi K^0$}}

\begin{figure}[hbt!]
\begin{center}
\epsfig{file=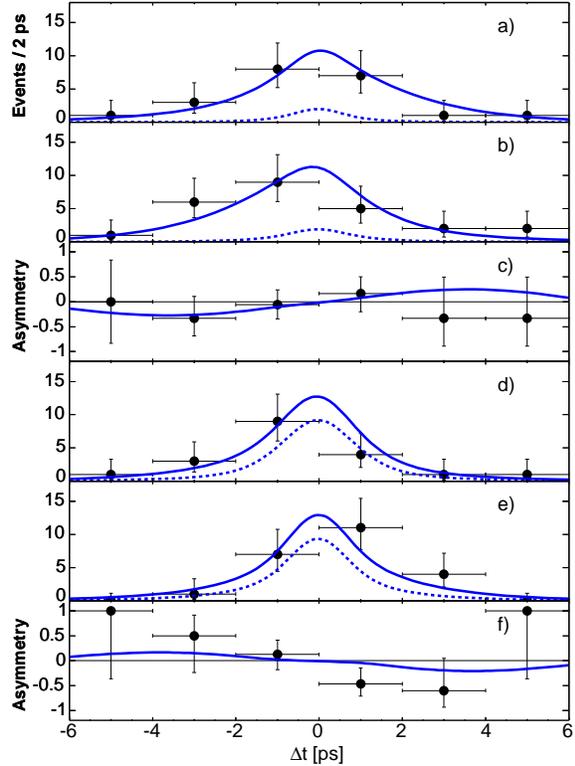, clip=, width=0.5\textwidth}
\caption{Plots of the $\Delta t$ distributions, for (a) $B^0$- and (b) 
$\bar{B}^0$-tagged $\phi K^0_S$ events, with plot c) showing the asymmetry. Plots d), e) and f)
show the corresponding plots for $\phi K^0_L$ candidates.}
\label{fig:BPhiKsAsymm}
\end{center}
\end{figure}

The mode $B \ra \phi K^0$ is reconstructed using both $K^0_S$ and $K^0_L$
decays, with $\phi \ra K^+ K^-$ and $K^0_S \ra \pi^+ \pi^-$. The two-kaon invariant
mass is required to be within 16~\mevcc of the nominal $\phi$ mass~\cite{pdg}, while
$K^0_L$ candidates are identified using information from the EMC and IFR.
To validate the analysis, the control sample $B^{\pm} \ra \phi K^{\pm}$ is used, in which
the measured \CP-asymmetry parameters are $S = 0.23 \pm 0.24$ and $C = -0.14 \pm 0.18$,
which are consistent with the SM expectation of no \CP\ violation. The results
for this mode presented in Table~\ref{tab:PenguinResults} are consistent with the SM.
Figure~\ref{fig:BPhiKsAsymm} shows the decay time distributions for this mode.

\subsection{\boldmath{$B^0 \ra K^+ K^- K^0_S$}}

The decay $B^0 \ra K^+ K^- K^0_S$ has the same final state as the previous mode,
except that events containing the $\phi \ra K^+ K^-$ resonance are removed. This sample
is several times larger than the sample of $\phi K^0_S$, and therefore provides
a more accurate way to measure the \CP-violating parameters for this final state.
The measured \CP-even fraction of this decay, equal to 
$2 \Gamma(B^+ \ra K^+ K^0_S K^0_S)/\Gamma(B^0 \ra K^+ K^- K^0)$, is $0.98 \pm 0.15 \pm 0.04$. 
This implies that the final state is \CP-even dominated. 
The values of $S$ and $C$ shown in Table~\ref{tab:PenguinResults}
are consistent with the SM, and setting $C$ to zero gives a value of
sin$2\beta$ of $0.57 \pm 0.26 \pm 0.04^{+0.17}_{-0.00}$, where the last error
represents the uncertainty on the \CP content. This result is consistent
with the world-average value of $0.73 \pm 0.05$~\cite{pdg}.

\subsection{\boldmath{$B^+ \ra K^+ K^0_S K^0_S$}}

In the SM, we expect that the decay rates for $B^+ \ra K^+ K^0_S K^0_S$
and $B^- \ra K^- K^0_S K^0_S$ to be equal, although contributions from physics
beyond the SM could give a non-zero direct \CP\ asymmetry. We
measure an asymmetry that is consistent with zero, as shown in 
Table~\ref{tab:PenguinResults}.

\subsection{\boldmath{$B^0 \ra f_0(980) K^0_S$}}

This mode is reconstructed with the decays $f_0(980) \ra \pi^+ \pi^-$ and
$K^0_S \ra \pi^+ \pi^-$. The invariant mass of the $f_0(980)$ resonance is required
to be between 0.86 and 1.10~\gevcc, and is parameterised as a relativistic
Breit-Wigner, with a measured mass of $(980.6 \pm 4.1 \pm 0.5 \pm 4.0)~\mevcc$
and width of $(43^{+12}_{-9} \pm 3 \pm 9)~\mevcc$, where the last errors are
uncertainties due to interference effects from other resonances
in the $B^0 \ra K^0_S \pi^+ \pi^-$ Dalitz plot. These values are in agreement
with previous measurements~\cite{pdg}. The results of $S$ and $C$ for this mode
(see Table~\ref{tab:PenguinResults}) are consistent with the SM
at the $1.7\sigma$ and $0.8\sigma$ levels, respectively.

\section{\boldmath{$B^0 \ra \rho^+ \rho^-$}}

The time-dependent \CP-violating
asymmetry in the decay $B^0 \ra \rho^+ \rho^-$ is related to the CKM
angle $\alpha$. If the decay proceeds only through tree diagrams, then the asymmetry
is directly related to $\alpha$. However, we can only measure 
an effective angle, $\alpha_{\rm{eff}}$, if there is pollution from gluonic penguins.
Recent measurements of the $B^+ \ra \rho^+ \rho^0$ branching fraction and upper limit
for $B^0 \ra \rho^0 \rho^0$~\cite{rhoBF} indicate that the penguin pollution is small,
which has been argued theoretically~\cite{Aleksan}.
It has also been found that the longitudinal polarisation dominates 
this decay~\cite{RhoLong}, which simplifies the \CP\ analysis.

The $\rho$ candidates are required to have an invariant mass
between 0.5 and 1.0~\gevcc, and combinatorial backgrounds are suppressed by applying
selection criteria to several event shape variables. An unbinned maximum likelihood fit
is performed to extract the following preliminary \CP-asymmetry results, assuming that the decay
has zero transverse polarisation: $S_{\rm{long}} = -0.19 \pm 0.33 \pm 0.11$ and
$C_{\rm{long}} = -0.23 \pm 0.24 \pm 0.14$, where the first errors are statistical and
the second errors are systematic uncertainties. The branching fraction for this mode
is measured to be $(30 \pm 4 \pm 5) \times 10^{-6}$.

\begin{figure}[hbt!]
\begin{center}
\epsfig{file=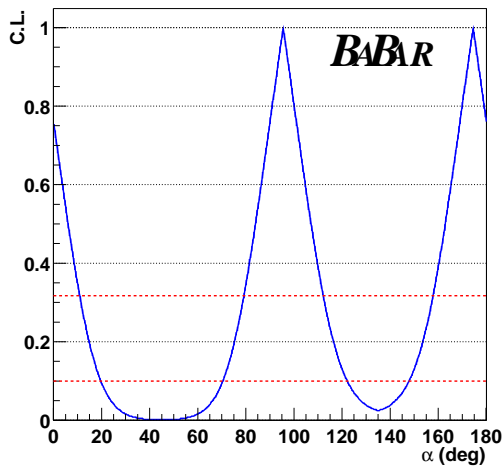, clip=, width=0.45\textwidth}
\caption{The value of $\alpha$ as a function of confidence level from the 
preliminary results of the isospin analysis of $B \ra \rho \rho$.
The dotted red lines represent the 10\% and 31.7\% confidence levels.}
\label{fig:rhoAlpha}
\end{center}
\end{figure}

The CKM angle $\alpha$ can be constrained by performing an isospin analysis
of $B \ra \rho\rho$~\cite{rhorhoAlpha}, which needs as input the amplitudes
of the \CP-even longitidunal polarisation of the $B$ meson decaying into $\rho^{\pm} \rho^0$,
$\rho^0\rho^0$ and $\rho^+ \rho^-$, and the measured values of $S_{\rm{long}}$ and $C_{\rm{long}}$
given above.
Assuming that isospin symmetry is valid, and that there are no significant non-resonant
or $I=1$ isospin contributions, the best CKM fit to the data gives the preliminary
result of $\alpha = (96 \pm 10 \pm 4 \pm 13)^{\degrees}$, where the last error
is the uncertainty from possible penguin pollution, which is estimated by using
the measured Grossman-Quinn bound~\cite{rhorhoAlpha}: $|\alpha_{\rm{eff}} - \alpha| < 13^{\degrees} (15.9^{\degrees})$ at
68.3\% (90\%) C.L. Figure~\ref{fig:rhoAlpha} shows the value of $\alpha$ as a function
of the confidence level.

\section{\boldmath{$B^0 \ra \phi K^{*0}(892)$}}

This mode is dominated by $b \ra s{\bar{s}}s$ penguin diagrams, like
the modes presented in Sec.~\ref{sec:PenguinModes}, and angular
correlation measurements and \CP-asymmetries are sensitive to contributions beyond
the SM. 

The decay rate of this channel depends on the helicity amplitudes  $A_{\lambda}$ of the vector
mesons, where $\lambda = 0$ or $\pm 1$~\cite{PhiKstTheory}. These
amplitudes can be expressed in terms of their \CP-even and \CP-odd equivalents:
$A_{||} = (A_{+1} + A_{-1})/\sqrt{2}$ and 
$A_{\perp} = (A_{+1} - A_{-1})/\sqrt{2}$. From this,
it follows that the longitudinal and transverse fractions are
$f_L = |A_0|^2/\sum |A_{\lambda}|^2$ and $f_{\perp} = |A_{\perp}|^2/\sum |A_{\lambda}|^2$,
respectively. The relative phases of the \CP-even and \CP-odd amplitudes are
$\phi_{||} = {\rm{arg}}(A_{||}/A_{0})$ and $\phi_{\perp} = {\rm{arg}}(A_{\perp}/A_0)$,
respectively. From the above, one can derive the vector triple-product asymmetries
${\cal{A}}_T^{||}$ and ${\cal{A}}_T^0$, which are sensitive to \CP-violation~\cite{TripleProd}:
\begin{equation}
{\cal{A}}^{||, 0}_T = \frac{1}{2}\left(\frac{ {\rm{Im}}(A_{\perp} A^*_{||,0})}{\sum|A_{\lambda}|^2} + 
\frac{ {\rm{Im}}(\bar{A}_{\perp} \bar{A}^*_{||,0})} {\sum|\bar{A}_{\lambda}|^2} \right),
\label{TripleProdEqn}
\end{equation}
where $\bar{A}_{\lambda}$ represents the conjugate helicity amplitude.

$B$ mesons are reconstructed by combining $\phi \ra K^+ K^-$
and $K^{*0} \ra K^+ \pi^-$ candidates. The invariant masses of the $\phi$ and $K^{*0}$
are required to be between 0.99 and 1.05~\gevcc, and between 1.13 and 1.73~\gevcc, respectively.
Continuum backgrounds are suppressed by using event shape variables.

\begin{figure}[hbt!]
\begin{center}
\epsfig{file=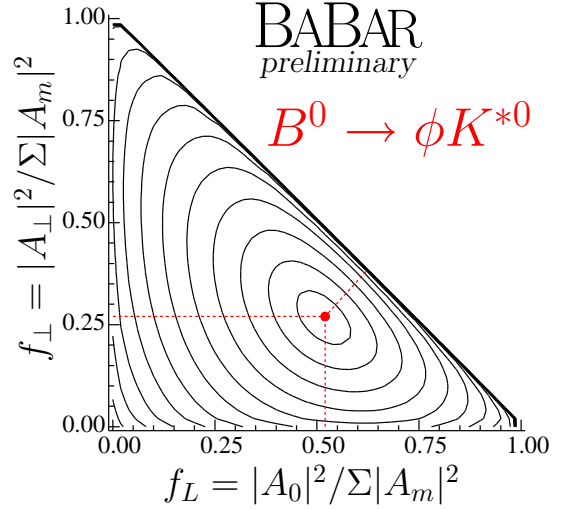, clip=, width=0.45\textwidth}
\caption{Plot of transverse (\CP-odd) versus longitudinal polarisation for $B^0 \ra \phi K^{*0}$,
showing likelihood function contours with $1\sigma$ intervals. The dot represents
the fit result.}
\label{fig:BPhiKstPol}
\end{center}
\end{figure}

Table~\ref{tab:phiKstresults} shows the preliminary results for this mode, using
an unbinned maximum likelihood fit to the data.
We observe non-zero contributions from all three helicity amplitudes $|A_0|$,
$|A_{||}|$ and $|A_{\perp}|$, with more than $5\sigma$ significance, as 
shown in Fig.~\ref{fig:BPhiKstPol}. The longitudinal
polarisation is essentially a factor of two less than that for $B \ra \rho \rho$~\cite{RhoLong},
which could be a hint of physics beyond the SM. However, this difference may
be related to long-distance effects from $c\bar{c}$ penguins~\cite{PhiKstLongTheory}.
There is no evidence for direct \CP\ violation.
\begin{table}[htb!]
\begin{center}
\caption{Preliminary results of the angular analysis of the decay $B^0 \ra \phi K^{*0}$. 
${\cal{B}}$ denotes the branching fraction, while ${\cal{A}}_{\CP}$, ${\cal{A}}^0_{\CP}$
and ${\cal{A}}^{\perp}_{\CP}$ denote the direct, longitudinal and transverse \CP-asymmetries, 
respectively. The \CP-even and \CP-odd phase differences are given by 
$\Delta \phi_{||} = \frac{1}{2}(\phi_{||}^+ - \phi_{||}^-)$ and
$\Delta \phi_{\perp} = \frac{1}{2}(\phi_{\perp}^+ - \phi_{\perp}^-)$, respectively, while
the triple product asymmetries are denoted by ${\cal{A}}^{||,0}_T$.}
\label{tab:phiKstresults}
\begin{tabular}{lc}
\hline
Variable & Result\\
\hline
${\cal{B}}$ & $(9.2 \pm 0.9 \pm 0.5) \times 10^{-6}$ \\
$f_{L}$ & $0.52 \pm 0.07 \pm 0.02$ \\
$f_{\perp}$ & $0.27 \pm 0.07 \pm 0.02$ \\
$\phi_{||}$ & $2.63^{+0.24}_{-0.23} \pm 0.04$ \\
$\phi_{\perp}$ & $2.71^{+0.22}_{-0.24} \pm 0.03$ \\
${\cal{A}}_{\CP}$ & $-0.12 \pm 0.10 \pm 0.03$ \\
${\cal{A}}^0_{\CP}$ & $-0.02 \pm 0.12 \pm 0.01$ \\
${\cal{A}}^{\perp}_{\CP}$ & $-0.10^{+0.25}_{-0.27} \pm 0.04$ \\
$\Delta \phi_{||}$ & $0.38^{+0.23}_{-0.22} \pm 0.03$ \\
$\Delta \phi_{\perp}$ & $0.30^{+0.24}_{-0.22} \pm 0.03$ \\
${\cal{A}}^{||}_T$ & $0.02 \pm 0.05 \pm 0.01$ \\
${\cal{A}}^0_T$ & $0.11 \pm 0.07 \pm 0.01$ \\
\hline
\end{tabular}
\end{center}
\end{table}

\section{Conclusions}

We have shown a selection of results from the \babar\ experiment 
based on up to $124~\invfb$ collected at the \FourS\ resonance.
The SM is consistent with the measurements presented here, although
there are hints of physics beyond the SM in $b \ra s$ penguin modes.
We can expect a more definite conclusion to this exciting prospect
in the near future.

\end{document}